Boquan Ren, Hongguang Wang, Victor O. Kompanets, Yaroslav V. Kartashov, Yongdong Li and Yiqi Zhang*

# Dark topological valley Hall edge solitons

**Abstract:** Topological edge solitons propagating along the edge of a photonic topological insulator are localized self-sustained hybrid states that are immune to defects/disorders due to the protection of the edge states stemming from the nontrivial topology of the system. Here, we predict that exceptionally robust dark valley Hall edge solitons may form at the domain walls between two honeycomb lattices with broken inversion symmetry. The underlying structure can be created with femtosecond laser inscription, it possesses a large bandgap where well-localized dark edge solitons form, and in contrast to systems with broken time-reversal symmetry, it does not require external magnetic fields or complex longitudinal waveguide modulations for the realization of the topological phase. We present the envelope equation allowing constructing dark valley Hall edge solitons analytically. Such solitons propagate without radiation into the bulk of the lattice and can circumvent sharp corners, which allows observing their persistent circulation along the closed triangular domain wall boundary. They survive over huge distances even in the presence of disorder in the underlying lattice. We also investigate interactions of closely located dark topological valley Hall edge solitons and show that they are repulsive and lead to the formation of two gray edge solitons, moving with different group velocities departing from group velocity of the linear edge state on which initial dark solitons were constructed. Our results illustrate that nonlinear valley Hall systems can support a rich variety of new self-sustained topological states and may inspire their investigation in other nonlinear systems, such as atomic vapors and polariton condensates.

## 1 Introduction

After more than a decade of development of topological photonics [1–4], it was recognized that nonlinear effects, such as self-action and parametric interactions, may fundamentally affect the evolution of excitations and change the very structure of the spectrum of topological systems, providing sometimes a very convenient knob for manipulation of the edge states and control of related localization and transport phenomena. Nonlinear topological photonics [5] turns into a rapidly expanding and fascinating discipline. Photonic systems are especially advantageous in this sense, because they frequently have strong nonlinearity [5] and, at the same time, allow the realization of topologically nontrivial potential landscapes [6, 7], including non-Hermitian ones [8, 9]. Representative examples of novel phenomena encountered in various (not necessary purely photonic) topological systems and based on nonlinear effects include topological insulator lasers [10–17], nonlinearity-induced topological phases [18–20] observed recently in waveguide arrays [21], bistability effects [22, 23], nonlinear tuning of the edge state energies [24], and topological edge solitons. The interplay between nonlinearity and topology was studied in both Hermitian and non-Hermitian one-dimensional (1D) photonic Su–Schrieffer–Heeger lattices [25–27].

Topological edge solitons appear as in-gap localized hybrid states, typically moving (in the two-dimensional [2D] or three-dimensional case) along the edge of the

*Corresponding author: Yiqi Zhang, Key Laboratory for Physical Electronics and Devices of the Ministry of Education & Shaanxi Key Lab of Information Photonic Technique, School of Electronic Science and Engineering, Xi'an Jiaotong University, Xi'an 710049, China, E-mail: zhangyiqi@xjtu.edu.cn. https://orcid.org/0000-0002-5715-2182

**Boquan Ren, Hongguang Wang and Yongdong Li,** Key Laboratory for Physical Electronics and Devices of the Ministry of Education & Shaanxi Key Lab of Information Photonic Technique, School of Electronic Science and Engineering, Xi'an Jiaotong University, Xi'an 710049, China, E-mail: boquanren@163.com (B. Ren), wanghg@xjtu.edu.cn (H. Wang), leyond@xjtu.edu.cn (Y. Li). https://orcid.org/0000-0002-0126-3131 (B. Ren). https://orcid.org/0000-0001-8994-4523 (H. Wang). https://orcid.org/0000-0001-8756-0438 (Y. Li)

**Victor O. Kompanets and Yaroslav V. Kartashov,** Institute of Spectroscopy, Russian Academy of Sciences, Troitsk, Moscow, 108840, Russia, E-mail: kompanetsvo@isan.troitsk.ru (V.O. Kompanets), yaroslav.kartashov@icfo.eu (Y.V. Kartashov). https://orcid.org/0000-0002-2500-9832 (V.O. Kompanets). https://orcid.org/0000-0001-8692-982X (Y.V. Kartashov) insulator with a velocity approximately set by the group velocity of the linear edge state, from which they bifurcate. On the one hand, edge solitons may pass defects without backscattering since they also enjoy topological protection, and, on the other hand, they maintain their envelopes (localization along the edge) due to the nonlinear self-action. In array-based photonic Floquet topological insulators, where time-reversal symmetry is broken by the artificial magnetic field due to the helicity of the waveguides, topologically closed currents in the bulk were first reported theoretically [28] and then observed experimentally [29]. In such Floquet systems, topological edge solitons were thoroughly investigated in both discrete [30–32] and continuous [33–36] models and observed in the anomalous regime in [37]. In polariton topological insulators, where time-reversal symmetry is broken by real magnetic fields in the presence of spin–orbit coupling, topological edge solitons were found in [38–41]. Besides this, a different type of Dirac [42] or Bragg topological solitons [43] may exist. Experimental progress in soliton observation in such systems was, however, severally limited due to strong material or radiative (due to waveguide helicity) losses typical for the above insulator types. Thus, Floquet edge solitons were reported only over one driving period [37]. Despite the observation of non-topological edge solitons [44], corner solitons in second-order insulators [45], and nonlinear modes in the SSH chains [25, 27], their long-living traveling counterparts in 2D insulators still require experimental exploration.

From the point of view of the experimental exploration of such solitons, photonic valley Hall insulators constructed on two detuned lattices (for demonstration of linear edge states in such systems see, for instance [46–53]) may provide a powerful alternative to Floquet or polaritonic platforms. The word "valley" is associated here with specific features (presence of the local extrema) of bands of corresponding systems: for example, when inversion symmetry of the underlying honeycomb lattice is broken by detuning of two constituent sublattices, the gap opens between former Dirac points and local extrema in two upper bands develop that are called valleys. Even though such valley Hall structures provide weaker topological protection, their advantage is the ease of fabrication and the fact that they may be implemented with straight waveguides [54] and, for this reason, will feature drastically reduced losses. Topological edge states in valley Hall systems are associated with broken inversion symmetry; they may form at the domain wall between honeycomb lattices with different detuning. Very recently valley Hall effect was proposed for nonlinear vortices nested in infinite modulated background [55]. At the same time, topological edge solitons in valley Hall systems are still open for exploration, to the best of our knowledge.

In this article, we predict that such solitons can form, develop their theory, and, for the first time to our knowledge, report on exceptionally robust dark valley Hall edge solitons, and analyze their interactions that may lead to the formation of pairs of topological gray solitons moving along the edge with different velocities, despite the fact that they are constructed on (nested in) the same edge state. Robustness of the dark valley Hall edge solitons is confirmed by their survival after a long propagation distance even in the presence of disorder in the lattice. Dark valley Hall edge solitons can be experimentally realized in laser-written waveguide arrays [6, 56], in optical lattices imprinted in a photorefractive crystal [54], or in atomic vapors [44].

## 2 Results and discussion

### 2.1 The model

The propagation of light along the $z$-axis in our system is governed by the dimensionless nonlinear Schrödinger equation for the light field amplitude $\psi$:

$$i\frac{\partial \psi}{\partial z} = -\frac{1}{2}\left(\frac{\partial^2}{\partial x^2} + \frac{\partial^2}{\partial y^2}\right)\psi - \mathscr{R}(x,y)\psi - |\psi|^2\psi, \qquad (1)$$

in which we assume focusing cubic nonlinearity of the material, and describe fabricated or optically induced refractive index profile by the function $\mathscr{R}(x,y) = \mathscr{R}_A(x,y) + \mathscr{R}_B(x,y)$, where two constituent sublattices are described by the functions $\mathscr{R}_{A,B}(x,y) = p_{A,B}\sum_{n,m}e^{-[(x-x_n)^2+(y-y_m)^2]/\sigma^2}$, where $(x_n, y_m)$ are the coordinates of the waveguides of corresponding sublattices in the honeycomb structure, and $\sigma = 0.5$ is the waveguide width. The waveguide depths for sublattices $A$ and $B$ forming honeycomb lattice are determined by $p_{A,B} = p \pm \delta$, where $p = 10.3$ and detuning $\delta = 0.55$. For the characteristic transverse scale of 10 μm corresponding to the dimensionless coordinate $x = 1$ and a wavelength of 800 nm the above-mentioned refractive index modulation depth corresponds to $\sim 1.1 \times 10^{-3}$. Such shallow arrays can be inscribed in fused silica using the femtosecond-laser writing technique [6, 56]. Numerical simulations demonstrate that the waveguides with these parameters are single-mode. The detuning $\delta$ breaks the inversion symmetry of the honeycomb lattice. The eigenmodes of such infinite lattice – Bloch waves – can be found from the linear counterpart of Eq. (1), by representing $\psi = \phi(x,y,k_{x,y})e^{ibz}$, where $\phi$ is the Bloch mode profile, $k_{x,y}$ are Bloch momenta, and $b$ is the propagation constant. Using the plane-wave expansion method, we obtained the band structure of the inversion-symmetry-broken honeycomb lattice with $p_A = p - \delta$ and $p_B = p + \delta$, two top bands of which are shown in Figure 1(a). Clearly, the Dirac cones between the upper and lower bands disappear and a wide gap opens up. The honeycomb lattice with inverted detuning, i.e. $p_A = p + \delta$ and $p_B = p - \delta$ exhibits the same band structure as in Figure 1(a); however, Berry curvature $\Omega = \nabla_\mathbf{k} \times [i \langle \phi_\mathbf{k} | \nabla_\mathbf{k} \phi_\mathbf{k} \rangle]$ [57] corresponding to the top or bottom band is now opposite. By joining two inversion-symmetry-broken honeycomb lattices with opposite detuning at $x < 0$ and $x \geq 0$, one creates a straight domain wall, where the effective refractive index is locally increased, as depicted in Figure 1(b) by the red ellipse. Composite honeycomb lattice is periodic along the $y$ axis, so the function $\mathscr{R}$ fulfills the condition: $\mathscr{R}(x, y) = \mathscr{R}(x, y + L)$ with $L = \sqrt{3}d$ being the $y$-period, and $d = 1.4$ being the separation between two nearest waveguides. We set outer boundaries along $x$ so far from the interface at $x = 0$ that they do not affect propagation along this interface. Linear eigenmode of the composite lattice is now written as $\psi = \phi(x, y)e^{ibz} = u(x, y)e^{ik_y y + ibz}$, where $u(x, y) = u(x, y + L)$, $k_y$ is the Bloch momentum in the first Brillouin zone (BZ) $-K/2 \leq k_y \leq K/2$ with $K = 2\pi/L$ being the width of the first BZ. The corresponding "projected" band structure $b(k_y)$ is shown in the gray plane in Figure 1(a), in which the black curves are the bulk states, green and cyan curves correspond to the edge states at far-away outer bearded boundaries of the lattice (we do not consider them), while the red curve is the topological valley Hall edge state that we will investigate here. The existence of the valley Hall edge states is mediated by the valley Hall effect [46–49]. Across the domain wall, the Berry curvature for a given valley changes its sign, so that the difference of the corresponding topological indices (valley Chern numbers) becomes ±1, which means that edge states should appear at the domain wall according to the bulk-edge correspondence principle [1, 2]. Two representative edge states are indicated by the red ($k_y = -0.476K$) and blue ($k_y = -0.3K$) dots on the red curve, and presented in Figure 1(c). As expected, the state at $k_y = -0.3K$ (blue dot) with $b$ deep in the gap is concentrated in the vicinity of the domain wall and only slightly penetrates into bulk. The edge state at $k_y = -0.476K$ [red dot in Figure 1(a)] is too close to the bulk band, therefore its localization is much weaker.

## 2.2 Nonlinear valley Hall edge states

To highlight the dispersion properties of the edge state, we show in the yellow plane of Figure 1(a) the first-order derivative $b' = \partial b / \partial k_y$ (solid curve) and the second-order derivative $b'' = \partial^2 b / \partial k_y^2$ (dashed curve). The first of them determines the group velocity of the edge state $v = -b'$, while $b''$ determines its dispersion rate. Since $b'' > 0$ in a wide range in the first BZ, one may assume that such a composite lattice may support dark valley Hall edge solitons localized along the domain wall. Their robustness, however, requires stability of the background far from the soliton center. Therefore, before seeking the dark valley Hall edge soliton, it is necessary to check the stability of the nonlinear periodic valley Hall edge states that would serve as such a background for dark solitons. Such nonlinear edge states $\psi(x, y, z) = u(x, y)e^{ibz + ik_y y}$ with $u(x, y) = u(x, y + L)$, bifurcating from linear ones can be found iteratively using the Newton method from the following equation obtained from Eq. (1):

$$bu = \frac{1}{2}\left(\frac{\partial^2}{\partial x^2} + \frac{\partial^2}{\partial y^2} + 2ik_y\frac{\partial}{\partial y} - k_y^2\right)u + \mathcal{R}(x,y)u + |u|^2 u, \quad (2)$$

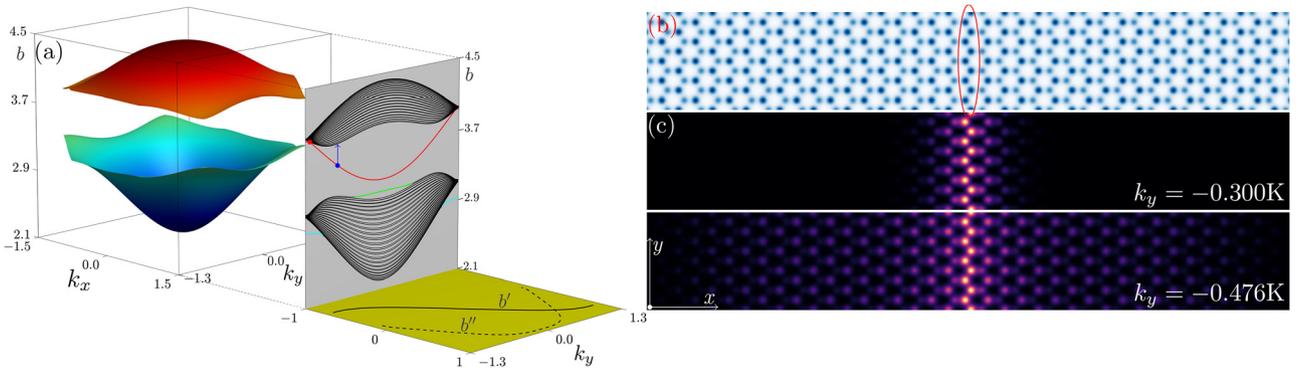

**Figure 1:** Band structures and linear valley Hall edge states.
(a) Band structure $b(k_x, k_y)$ of an inversion-symmetry broken honeycomb lattice. In the gray plane, the band structure of the composite honeycomb lattice with boundaries along $x$ is displayed, in which the red curve is the topological valley Hall edge state, black curves are bulk states, while green and cyan curves correspond to trivial edge states on the outer boundaries of the composite lattice. In the yellow plane, the first-order $b'$ (solid curve) and second-order $b''$ (dashed curve) derivatives of the propagation constant of the edge state are shown. (b) Honeycomb lattice with a domain wall (indicated by the red ellipse) created by the opposite detuning at $x < 0$ and $x \geq 0$. (c) Profiles of the edge states corresponding to the blue ($k_y = -0.3K$) and red ($k_y = -0.476K$) dots in (a).

We choose linear valley Hall edge state at $k_y = -0.3K$ as the bifurcation point and find nonlinear valley Hall edge states with the same momentum and propagation constant $b$ belonging to the topological gap.

The peak amplitude $a = \max|\psi|$ and power $P = \int_{-\infty}^{+\infty} dx \int_0^L dy |\psi|^2$, concentrated within one $y$-period, for the nonlinear valley Hall edge state family at this momentum $k_y$ are shown in Figure 2(a), where the topological gap is shown white, while the bulk band is shown gray. We also display the profiles of the selected nonlinear valley Hall edge states in Figure 2(b), with their nonlinear propagation constants indicated in the right-bottom corner in each panel. One can see that the nonlinear edge state bifurcates from the linear one as its peak amplitude $a$ increases. When propagation constant $b$ of the nonlinear edge state approaches the upper border of the topological gap, the nonlinear valley Hall edge state becomes delocalized due to coupling with bulk modes: this tendency for delocalization is visible already for the state with $b = 3.825$ in Figure 2(b). To check the stability of the nonlinear valley Hall edge states, we superimpose on them a random noise with maximal amplitude up to $0.05a$ and propagate them over a very long distance far exceeding any experimentally available sample length. The example of propagation for weakly perturbed nonlinear edge state with $b = 3.604$ corresponding to the green dot in Figure 2(a) is shown in Figure 2(c). Its peak amplitude remains practically unchanged up to $z = 10^4$ (it only slightly oscillates upon propagation), so we conclude that such small-amplitude states are sufficiently robust. Instabilities may develop, but for states with substantially larger amplitudes $a \sim 0.5$ and for propagation constants close to the gap edge.

## 2.3 Dark valley Hall edge solitons

To obtain dark valley Hall edge solitons we assume that such solitons bifurcate from linear in-gap edge states $\psi_k$, and look for the solution of Eq. (1) in the form $\psi = A(Y,z)\phi_k e^{ibz}$, where $A$ is the slowly varying envelope, $\phi_k = u e^{ik_y y}$ is the linear edge state profile taken at selected Bloch momentum, and $Y = y - v_s z$ is the coordinate in the frame moving with group velocity $v_s = -b'$ at $k = k_y$. Using multiple-scale expansion and following derivation procedure described in [36] for continuous topological systems (since now the lattice does not change with $z$, no averaging in the propagation direction is required), one can show that the evolution of the envelope $A(Y,z)$ is governed by the nonlinear Schrödinger equation

$$i\frac{\partial A}{\partial z} = \frac{b''}{2}\frac{\partial^2 A}{\partial Y^2} - \chi|A|^2 A. \quad (3)$$

Here, $\chi = \int_{-\infty}^{+\infty} dx \int_0^L dy |\phi_k|^4$ is the effective nonlinear coefficient defined by the integral from the corresponding linear edge, while integration is performed over one $y$-period of the lattice. Dark solitons of Eq. (3) have the form $A = (b_{\rm nl}/\chi)^{1/2}\tanh[(b_{\rm nl}/b'')^{1/2}Y]e^{ib_{\rm nl}z}$, where $b_{\rm nl}$ is the nonlinearity-induced propagation constant shift that should be sufficiently small to ensure that the envelope is sufficiently broad so that it weakly changes on one $y$-period.

An example of the dark valley Hall edge soliton with $b_{\rm nl} = 0.0014$ prepared by superimposing the above-mentioned analytical envelope $A$ on the linear edge state $\psi_k$ is shown in Figure 3(a). To satisfy the periodicity of the field on our very extended (210 $y$-periods), but finite integration window, we nested two dark solitons in the same envelope. Propagation of this state in full Eq. (1) demonstrates that the soliton maintains its shape upon propagation along the domain wall. Selected profiles of propagating dark soliton are displayed in Figure 3(a) for distances indicated in the left-bottom corner of each panel. One finds that the profile at $z = 6000$ is practically the same as that at $z = 0$ except for overall displacement, i.e. solution of Eq. (3) provides a very accurate expression for the envelope. Remarkably, the soliton background also remains stable upon evolution, in agreement with stability results for low-amplitude nonlinear edge states. If the nonlinearity in Eq. (1) is omitted, the same input state exhibits considerable diffraction and broadening of two dark notches, as shown at $z = 6000$ in Figure 3(b). Since initial dark soliton is prepared by superimposing the envelope on the linear Bloch mode, one can calculate the projection of the field $\psi(x,y)$ on corresponding linear Bloch mode: $c(z) = \int_{-\infty}^{+\infty} dx \int_{mL-L/2}^{mL+L/2} \phi_{k_y}^*(x,y,z)\psi(x,y,z)dy$, where $m \in \mathbb{Z}$ defines the $y$-period, on which projection is calculated. These projections allow to explicitly track all deformations of soliton profile in the course of its propagation. In Figure 3(c), the projections (red dots) corresponding to the field from Figure 3(a) and 3(b) are displayed together with the initial envelope function $A$ (black line). Notice almost ideal conservation of soliton shape in the nonlinear case and strong broadening in the linear medium. Similar results are obtained for other momentum $k_y$ and sufficiently small $b_{nl}$ values.

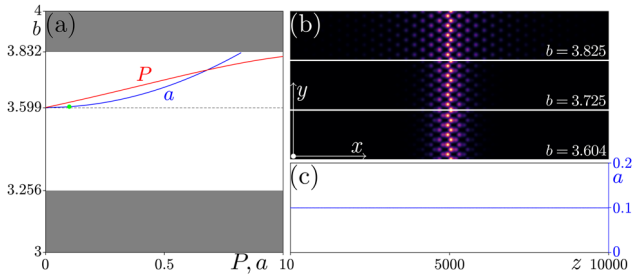

**Figure 2:** Nonlinear valley Hall edge state family.
(a) Peak amplitude $a$ (blue curve) and power $P$ on one $y$-period (red curve) of the nonlinear valley Hall edge state family. The green dot represents the mode with $b = 3.604$, whose peak amplitude $a \approx 0.1$. The gray regions represent bulk bands at $k_y = -0.3K$. The dashed line shows the propagation constant of the linear edge state. (b) Profiles of selected nonlinear valley Hall edge states at different values of $b$. (c) The peak amplitude of the perturbed nonlinear edge state with $b = 3.604$ versus propagation distance $z$.

To further illustrate the exceptional robustness of the dark valley Hall edge solitons, we introduce perturbations into the underlying lattice structure that is now described by the function $\mathcal{R}_{A,B}(x,y) = \sum_{n,m}(p_{A,B} + q_{n,m})e^{-[(x-x_n)^2+(y-y_m)^2]/\sigma^2}$, where $q_{n,m}$ is a random number uniformly distributed within the segment [–0.05, 0.05]. Importantly, this disorder is a small-scale one and it can potentially induce inter-valley scattering. However, propagation of dark solitons in such perturbed structures reveals that they maintain their profiles and internal structure even after very long propagation distances, showing obvious resistance against inter-valley scattering. Representative propagation dynamics of dark soliton corresponding to parameters of Figure 3 in disordered structure is illustrated in Figure 4(a), while corresponding projections are shown in Figure 4(b). One finds that not only the profile but also the velocity of motion of the dark valley Hall edge soliton are practically not affected by the disorder in the lattice. This indicates that such structures should be readily observable in experiments, where small imperfections of the lattice profile are unavoidable upon its fabrication/induction. At the same time, it should be mentioned that if one introduces strong perturbation into lattice by removing one side of the lattice at the domain wall, dark soliton may be reflected, which is typical for the valley Hall systems.

## 2.4 Topological protection of the dark valley Hall edge soliton

As mentioned above, one of the most remarkable properties of the topological edge solitons making them attractive for potential practical applications is their topological protection. This property is present in the nonlinear valley Hall system too. To illustrate this, we designed a domain wall with a closed triangular shape (see Supplementary material for the honeycomb lattice with a triangular domain wall) and launched into this structure a localized edge state (occupying only one side of the triangle) with two nested dark solitons to check whether they can circumvent the sharp corners of the domain wall. In Figure 5, we show the isosurface plot illustrating how two dark solitons (white stripes) move at a constant velocity from one side of the triangle to another side while maintaining good localization. It is obvious that dark solitons can circumvent sharp corners without radiating energy into bulk. This is also seen from intensity distributions shown at different distances $z$ in the same figure. After one complete loop, the solitons return to the initial side of the triangle. The small residual field remaining at two other sides is due to the fact that the initial beam does not cover the entire domain wall and spreads during propagation.

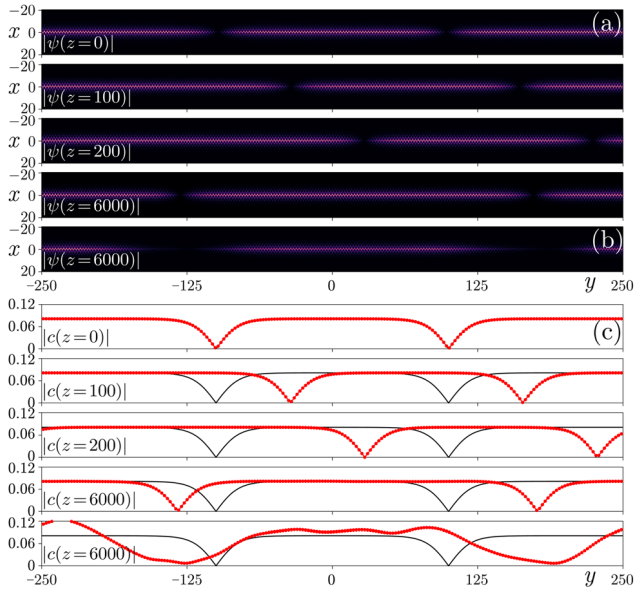

**Figure 3:** Dark topological edge soliton.
(a) Selected profiles illustrating nonlinear propagation of dark valley Hall edge soliton with $b_{nl} = 0.0014$ (distances are shown in the left-bottom corner of each panel). (b) Output field distribution at $z = 6000$ in a linear medium, for the same input as in panel (a). (c) Theoretical envelopes (solid curves) and amplitude projections (curves with dots) at $k_y = -0.3K$. For the convenience of presentation $y$-axis is made horizontal in this plot.

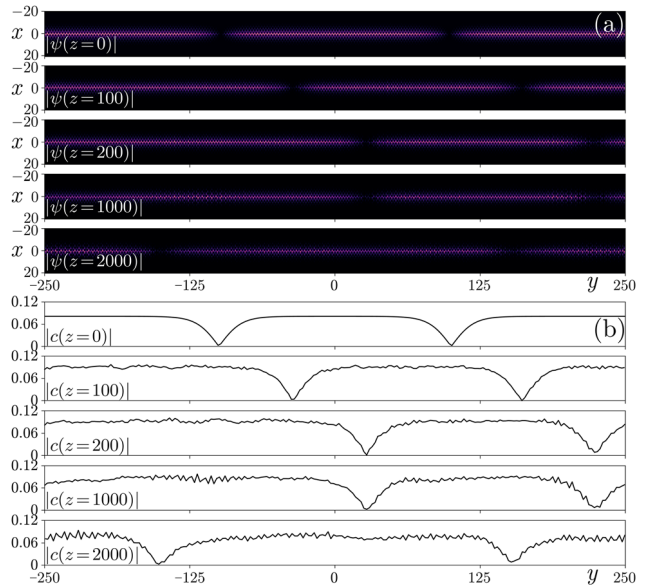

**Figure 4:** Profiles (a) and projections (b) of the dark valley Hall edge soliton at selective distances in the lattice with weak random disorders.

## 2.5 Repulsive interaction of valley Hall edge solitons

The possibility to nest several dark valley Hall solitons in the same edge state opens a unique opportunity to study their interactions in a topological insulator. The strength of such interactions is expected to depend on the initial separation between dark solitons, hence to study them we used the following envelope function at $z = 0$: $A_{z=0} = (b_{nl}/\chi)^{1/2}\tanh\left[(b_{nl}/b'')^{1/2}y_1\right]\tanh\left[(b_{nl}/b'')^{1/2}y_2\right]$ where coordinates of soliton centers are $y_1 = y - \Delta/2$, $y_2 = y + \Delta/2$, and $\Delta$ describes initial soliton separation. We found that interactions between topological dark solitons are repulsive and lead to increasing separation between them along the edge. This means that such solitons that initially move along the y-axis with equal velocities $v_s = -b'$ will acquire different asymptotic velocities (when separation becomes so large, that their interaction is negligible) that should depart from the group velocity $v_s$ of the linear edge state, in which dark solitons are nested. To calculate asymptotic velocities of the dark solitons we traced the positions of two minima in projection $c$ of the total field on Bloch wave $\phi_k$ up to sufficiently long distance $z \sim 4000$. This allowed us to calculate these velocities as a function of initial separation $\Delta$, as shown in Figure 6(a). The red curve in this plot corresponds to dark soliton initially located at $y = +\Delta/2$ that accelerates after the interaction, while the blue curve corresponds to soliton initially located at $y = -\Delta/2$ that slows down. The dashed line indicates the velocity $v_s \approx 0.6396$ of a single dark soliton. Clearly, smaller separation leading to stronger interaction produces larger velocity difference, while for large enough separations two solitons do not feel each other and their velocities approach that of the isolated state.

We also directly compared the outcome of dark soliton interactions in the full 2D topological valley Hall system and in reduced Eq. (3). To this end, we modeled the interaction of two 1D dark solitons in Eq. (3) with the same parameters that were used for the construction of the envelopes for 2D states, for various initial separations $\Delta$ between them, and for the same propagation distances. Corresponding asymptotic velocities acquired by two 1D dark solitons are shown in Figure 6(b), in which for convenience we show $V = v - v_s$. It is clear that variation in velocity due to interaction for 2D dark valley Hall edge solitons in Eq. (1) is very close to variation of velocities of 1D solitons, whose interaction is governed by Eq. (3), that again illustrates the validity of the latter equation for description of envelope evolution. In addition, we would like to note that the background which supports the dark solitons in the full 2D system is modulationally stable and for this reason, it allows nesting in its exceptionally robust dark states. An example of the interaction dynamics (in the form of dependence of projection $c$ on distance $z$) is presented in Figure 6(c) for $\Delta = 25$. Because the displacement of solitons is considerable and they may traverse large but finite integration window several times, we use numbers 1 and 2 to distinguish two solitons. One can see that not only does separation between two solitons increases indicating on their different asymptotic velocities but also solitons gradually become gray. This is especially obvious from the comparison of projections at $z = 0$ and $z = 4000$ in Figure 6(d). Therefore, the valley Hall system supports not only dark states remaining quiescent in the coordinate frame $Y = y − v_s z$ moving with the group velocity $v_s = −b'$ of the linear edge state with selected momentum $k$ but also gray states nested in the same Bloch wave, but moving with respect to it at nonzero velocity, obviously determining soliton grayness.

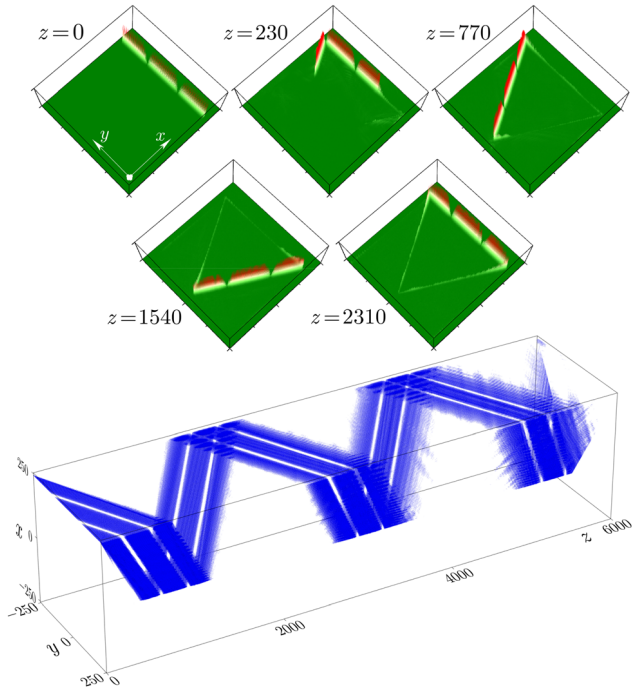

**Figure 5:** Topological protection of the dark valley Hall edge soliton. Isosurface intensity plot illustrating propagation of the dark valley Hall edge soliton along triangular domain wall, with the two white stripes corresponding to dark solitons. Field modulus distributions at different distances are also shown in the top five panels. Solitons corresponds to $k_y = -0.3K$ and $b_{nl} = 0.0014$.

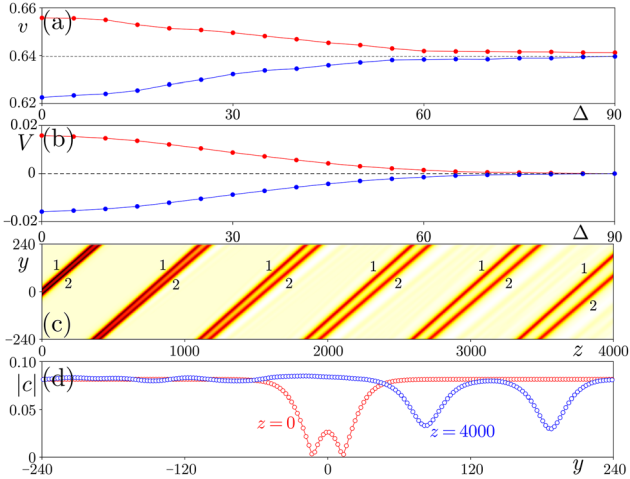

**Figure 6:** Repulsive interaction of two dark valley Hall edge solitons. (a) Asymptotic velocities of two interacting dark solitons versus initial separation Δ between them, based on Eq. (1). (b) Asymptotic velocities of two interacting one-dimensional dark solitons governed by the envelope Eq. (3). (c) Evolution of projections $c$ with distance for initial separation of Δ = 25. (c) Projections $c$ at $z = 0$ (red curve) and $z = 4000$ (blue curve).

## 3 Conclusions

To summarize, we have reported on robust dark valley Hall edge solitons at the domain wall between two honeycomb lattices with broken inversion symmetry. Such solitons can stably propagate along the straight domain wall and circumvent the sharp corners without reflection or radiation into the bulk. Repulsive interaction between two close dark valley Hall edge solitons change their velocities and transform them into gray edge solitons. Our results uncover reach possibilities for the investigation of nonlinear effects in topological systems with broken spatial symmetries that do not require magnetic fields or complex temporal driving for the onset of topological phases.

**Author contributions:** All the authors have accepted responsibility for the entire content of this submitted manuscript and approved submission.
**Research funding:** This work was supported by the National Natural Science Foundation of China (Nos. 12074308 and U1537210), the Russian Science Foundation (project No. 21-12-00096), and the Fundamental Research Funds for the Central Universities (No. xzy012019038).
**Conflict of interest statement:** The authors declare no conflicts of interest regarding this article.

# Supplementary Materials:
# Dark Topological Valley Hall Edge Solitons

Boquan Ren[1], Hongguang Wang[1], Victor O. Kompanets[2], Yaroslav V. Kartashov[2], Yongdong Li[1], and Yiqi Zhang[1,*]

[1]Key Laboratory for Physical Electronics and Devices of the Ministry of Education & Shaanxi Key Lab of Information Photonic Technique, School of Electronic Science and Engineering, Xi'an Jiaotong University, Xi'an 710049, China

[2]Institute of Spectroscopy, Russian Academy of Sciences, Troitsk, Moscow, 108840, Russia

[*]Corresponding authors: zhangyiqi@xjtu.edu.cn

**Topological protection of the dark valley Hall edge soliton**

The composite honeycomb lattice with a closed triangular domain wall created in it is displayed in Fig. S1. One can see that the domain wall is slightly darker than surrounding lattice regions. To depict the domain wall more clearly, we magnify its three corners (as indicated by three red boxes with numbers 1, 2 and 3) in the bottom three panels in Fig. S1. The structure of the domain wall around corners is highlighted by the red ellipses.

To obtain dynamics shown in Fig. 4 of the main text, we launched a pair of dark valley Hall edge solitons nested in common edge state at $k_y = -0.3K$ on the edge connecting corners 2 and 3 of the domain wall. As shown in Fig. 4 in the main text, this pair of dark valley Hall edge solitons will propagate along the domain wall in the counter-clockwise direction at a constant velocity and circumvent the corners of this structure without radiating energy into the bulk.

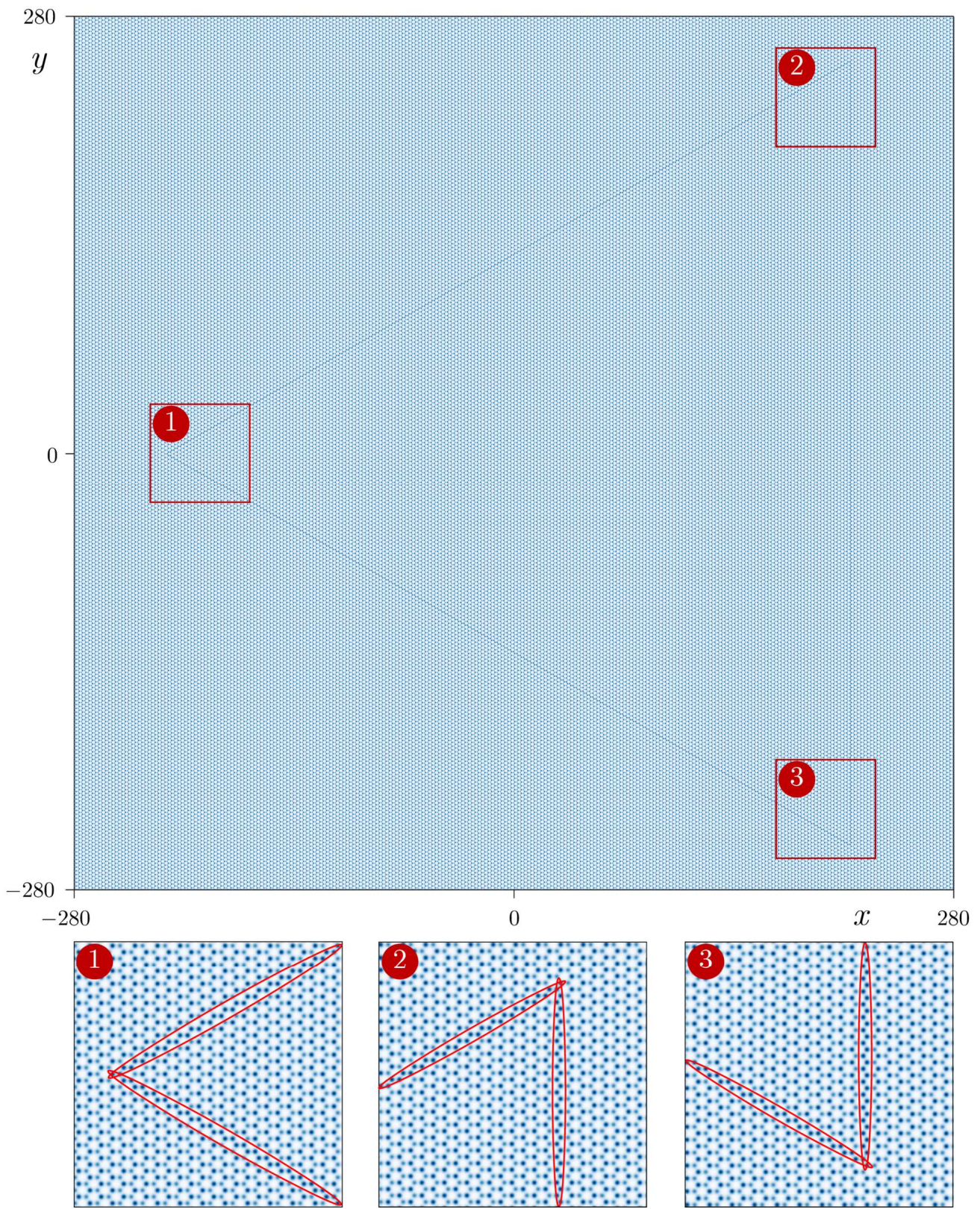

Fig. S1. Triangular domain wall created in the composite honeycomb lattice. Three corners of this domain wall are magnified in the bottom panels. In these panels, the position of the domain wall is indicated by the red ellipses.